\documentclass{PoS}

\title{{\footnotesize\vspace*{-2.cm}\hspace*{11.5cm} CERN-PH-TH/2012-307}\vspace*{2.cm}\\Implications of LHC Higgs and SUSY searches\\ for MSSM}

\ShortTitle{Implications of LHC Higgs and SUSY searches for MSSM}

\author{\speaker{Farvah Mahmoudi}\\
	CERN Theory Division, CH-1211 Geneva 23, Switzerland\\
        Clermont Universit\'e, Universit\'e Blaise Pascal, CNRS/IN2P3, LPC, BP 10448, F-63000 Clermont-Ferrand, France\\
        E-mail: \email{mahmoudi@in2p3.fr}}

\author{Alexandre Arbey\\
	Centre de Recherche Astrophysique de Lyon, Observatoire de Lyon, Saint-Genis Laval Cedex, F-69561, France; CNRS, UMR
	5574; Ecole Normale Sup\'erieure de Lyon, Lyon, France;
	Universit\'e de Lyon, France; Universit\'e Lyon 1, 
	F-69622~Villeurbanne Cedex, France\\
	CERN Theory Division, CH-1211 Geneva 23, Switzerland\\
E-mail: \email{alexandre.arbey@ens-lyon.fr}}

\author{Marco Battaglia\\
	Santa Cruz Institute of Particle Physics, University of California, Santa Cruz,
	CA 95064, USA; 
	Lawrence Berkeley National Laboratory, Berkeley, CA 94720, USA\\
	CERN, CH-1211 Geneva 23, Switzerland\\
E-mail: \email{marco.battaglia@cern.ch}}

\author{Abdelhak Djouadi\\
	Laboratoire de Physique Th\'eorique, Universit\'e Paris XI and CNRS, F-91405 Orsay, France;
	CERN Theory Division, CH-1211 Geneva 23, Switzerland\\
E-mail: \email{abdelhak.djouadi@cern.ch}}

\abstract{The implications of the LHC SUSY searches as well as the discovery of a new bosonic state compatible with the lightest Higgs boson will be discussed in the context of constrained and general MSSM scenarios. Exploring the MSSM through the Higgs sector is an alternative and complementary path to direct searches, and tight constraints on the MSSM parameter space can be obtained.  
}

\FullConference{36th International Conference on High Energy Physics\\
         4-11 July 2012\\
         Melbourne, Australia}

\begin{document}

\section{Introduction}
The search for Higgs and supersymmetry (SUSY) are the main focus of ATLAS and CMS experiments. Although before the start of the LHC the expectation for an early discovery of supersymmetric partners of the Standard Model (SM) particles was very high (mainly driven by the studies in the constrained SUSY scenarios), no SUSY particle has been observed yet. On the other hand, ATLAS and CMS collaborations have reported the discovery of a new bosonic state with a mass of around 126 GeV, compatible with the SM-Higgs \cite{ATLAS:2012gk,CMS:2012gu}. These results have significant implications for the Minimal Supersymmetric Standard Model (MSSM).
In the following, we discuss the consequences of the latest SUSY and Higgs search results in the context of the MSSM.

\section{Implication of SUSY searches}

To study the implication of the LHC SUSY searches we consider the unconstrained phenomenological MSSM (pMSSM) with 19 parameters \cite{Djouadi:1998di}. Most of the previous studies considered the highly constrained models with a small number of free parameters. However, these models are not representative of a generic MSSM scenario where the particle mass parameters are independent. As we will see below, the results can be very different in such generic scenarios. 

To explore the pMSSM, we perform a flat scan over the parameters in the ranges given in Table~\ref{tab:paramSUSY}. %
The particle spectra are generated for more than 100M points using {\tt SOFTSUSY} \cite{softsusy} and {\tt SUSPECT} \cite{suspect}. We impose the SUSY and Higgs mass limits from LEP and Tevatron as described in \cite{Arbey:2011aa}. The flavour observables, muon anomalous magnetic moment and relic density are computed with {\tt SuperIso Relic} \cite{superiso}, and we apply the constraints given in Table~\ref{tab:constraints}. We do not discuss here the consequences of the dark matter direct detection results which are discussed thoroughly in \cite{Arbey:2012na}.
\begin{table}[b!]
\begin{center}
\begin{tabular}{|c|c||c|c|}
\hline
~~~~Parameter~~~~ & ~~~~~~~~~~Range~~~~~~~~~~&~~~~Parameter~~~~ & ~~~~~~~~~~Range~~~~~~~~~~\\
\hline\hline
$\tan\beta$ & [1, 60]&$M_{\tilde{e}_L}=M_{\tilde{\mu}_L}$ & [50, 2500]\\
\hline
$M_A$ & [50, 2000]&$M_{\tilde{e}_R}=M_{\tilde{\mu}_R}$ & [50, 2500]\\
\hline
$M_1$ & [-2500, 2500]&$M_{\tilde{\tau}_L}$ & [50, 2500]\\
\hline
$M_2$ & [-2500, 2500]&$M_{\tilde{\tau}_R}$ & [50, 2500]\\
\hline
$M_3$ & [50, 2500]&$M_{\tilde{q}_{1L}}=M_{\tilde{q}_{2L}}$ & [50, 2500]\\
\hline
$A_d=A_s=A_b$ & [-10000, 10000]&$M_{\tilde{q}_{3L}}$ & [50, 2500]\\
\hline
$A_u=A_c=A_t$ & [-10000, 10000]&$M_{\tilde{u}_R}=M_{\tilde{c}_R}$ & [50, 2500]\\
\hline
$A_e=A_\mu=A_\tau$ & [-10000, 10000]&$M_{\tilde{t}_R}$ & [50, 2500]\\
\hline
$\mu$ & [-1000, 2000]&$M_{\tilde{d}_R}=M_{\tilde{s}_R}$ & [50, 2500]\\
\hline
&&$M_{\tilde{b}_R}$ & [50, 2500]\\
\hline
\end{tabular}
 \end{center}
\caption{SUSY parameter ranges (in GeV when applicable).\label{tab:paramSUSY}}
\end{table}%
To evaluate the consequences of the SUSY searches, we compute the supersymmetric particle decay rates with {\tt SDECAY} \cite{Muhlleitner:2003vg} and we use {\tt PYTHIA 6} \cite{Sjostrand:2006za} for event generation of inclusive SUSY production in $pp$ interactions. The generated events are then passed through fast detector simulation using {\tt Delphes} \cite{Ovyn:2009tx}. 
The Higgs decay rates are computed with {\tt HDECAY} \cite{Djouadi:1997yw} and the gluon fusion and VBF cross sections of the lightest CP-even Higgs with {\tt HIGLU} \cite{Spira:1995mt} and {\tt FeynHiggs} \cite{Heinemeyer:1998yj}. More details can be found in \cite{Arbey:2011aa,Arbey:2011un}.

\begin{table}
\begin{center}
\begin{tabular}{|c|}
\hline
$2.16 \times 10^{-4} < \mbox{BR}(B \to X_s \gamma) < 4.93 \times 10^{-4}$\\
\hline
$\mbox{BR}(B_s \to \mu^+ \mu^-) < 5.0 \times 10^{-9}$\\
\hline
$0.56 < \mbox{R}(B \to \tau \nu) < 2.70$\\
\hline
$4.7 \times 10^{-2} < \mbox{BR}(D_s \to \tau \nu ) < 6.1 \times 10^{-2}$\\
\hline
$2.9 \times 10^{-3} < \mbox{BR}(B \to D^0 \tau \nu) < 14.2 \times 10^{-3}$\\
\hline
$0.985 < \mbox{R}_{\mu23}(K \to \mu \nu)  < 1.013$\\
\hline
$-2.4 \times 10^{-9} < \delta a_\mu < 4.5 \times 10^{-9}$\\
\hline
$10^{-4} < \Omega_\chi h^2 < 0.155$\\
\hline
\end{tabular}
\end{center}
\caption{Constraints applied in our pMSSM analysis. The points passing all the constrained are called ``accepted points''.\label{tab:constraints}}
\end{table}%

\begin{figure}[t!]
\begin{center}
\vspace*{0.5cm}
\hspace*{-0.3cm}\includegraphics[width=0.35\textwidth]{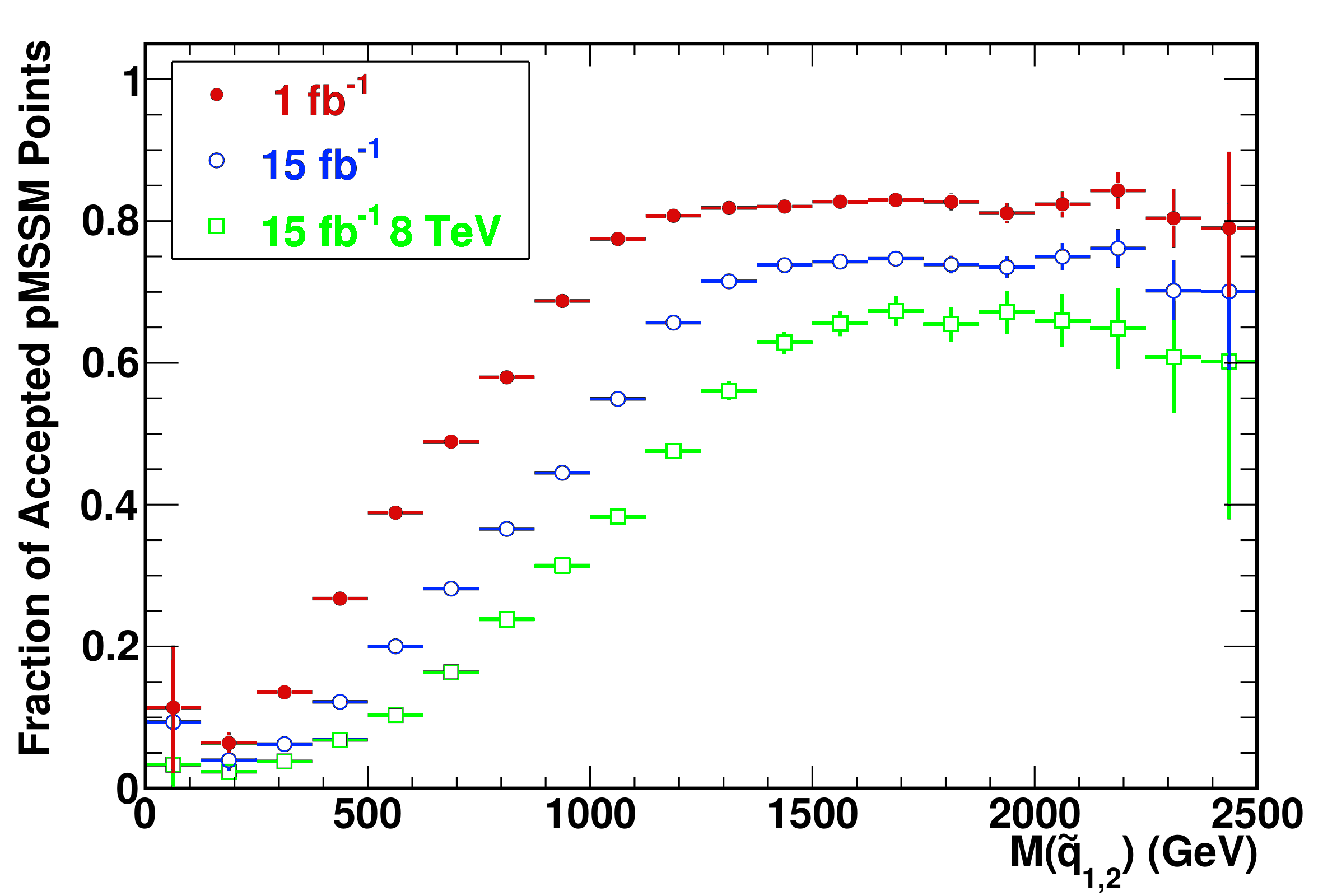}%
\includegraphics[width=0.35\textwidth]{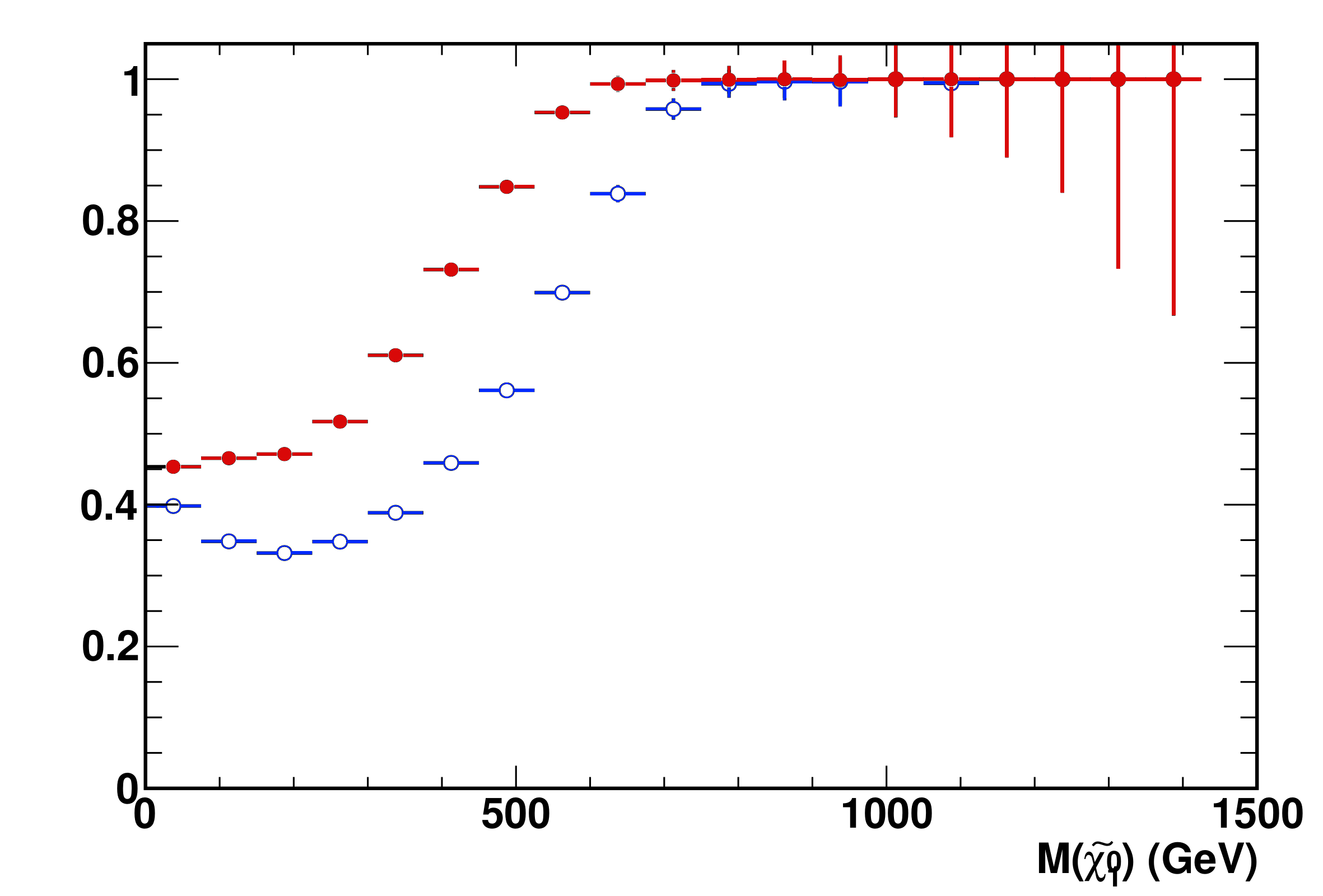}%
\includegraphics[width=0.35\textwidth]{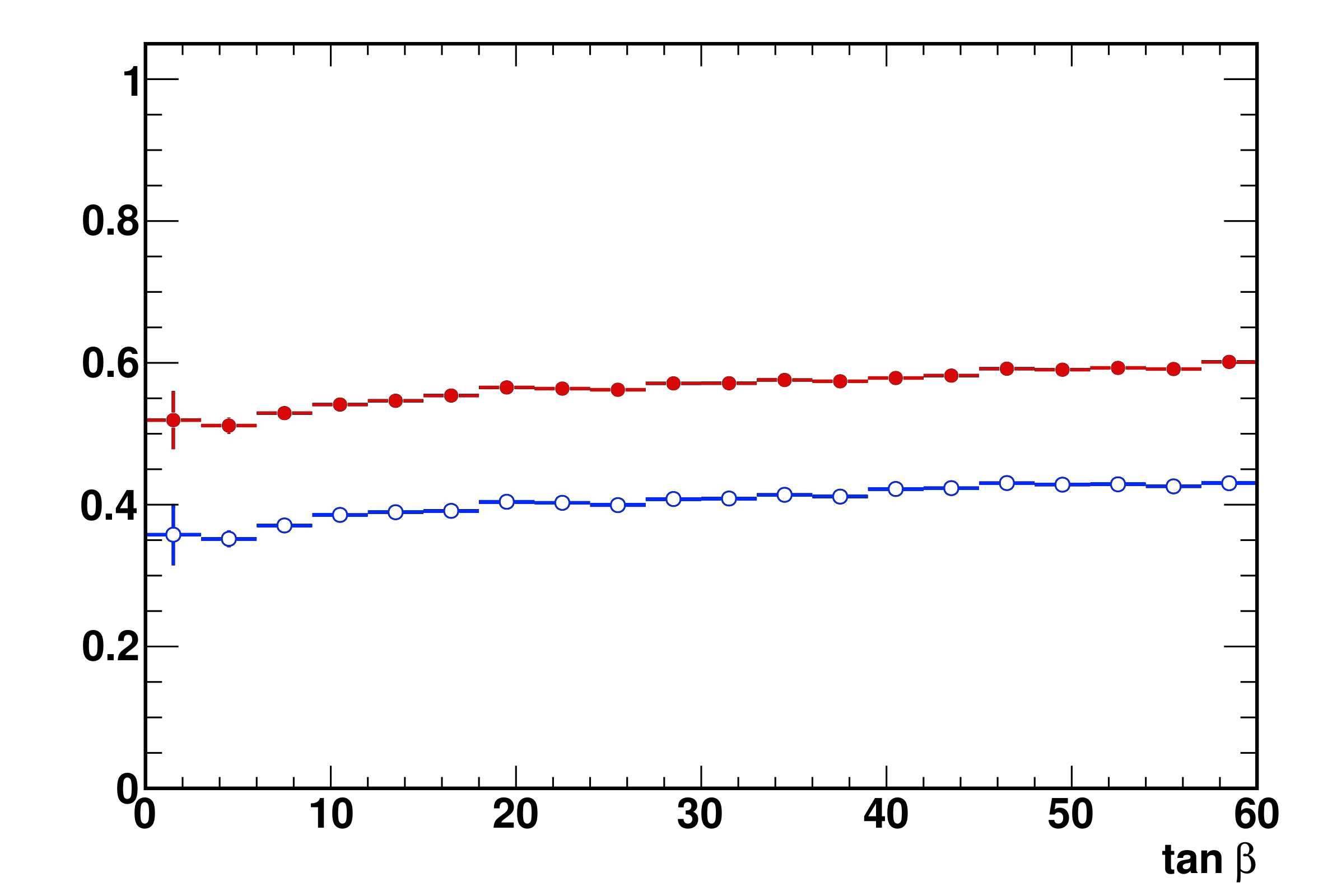}%
\end{center}
\caption{Fraction of accepted pMSSM points with $M_{h}>$ 111~GeV, not excluded by the SUSY searches with 1~fb$^{-1}$ of 7~TeV data (red), and by a projection of 15~fb$^{-1}$ at 7~TeV (blue) and 15~fb$^{-1}$ at 8~TeV (green), as functions of the masses of the lightest squark of the first two generations (left panel), the lightest neutralino $\tilde \chi^0_1$ (central panel) and $\tan \beta$ (right panel).}
\label{fig:susysearches}
\end{figure}

We consider the consequences of the SUSY searches in all hadronic events with $\alpha_T$ \cite{11-003}, in same-sign isolated dilepton events with jets \cite{11-010} and missing energy and in opposite-sign dilepton events with missing transverse energy \cite{11-011} in the CMS detector at 7~TeV with 1~fb$^{-1}$ of data, and extrapolate to the 8 TeV run with 15~fb$^{-1}$ of data. 
In Fig.~\ref{fig:susysearches} we show the consequences on the masses of the lightest squark of the first two generations, the lightest neutralino and $\tan \beta$, where the distribution of the points compatible with SUSY searches with 1~fb$^{-1}$ at 7~TeV and the projections for 15~fb$^{-1}$ at 7 and 8~TeV data are displayed.
We notice that with 15~fb$^{-1}$ of data at 8~TeV, more than 30\% of the points with squark masses below 1~TeV will still be allowed. The spectrum of the weakly interacting particles will be even less affected as there is basically no sensitivity for neutralino masses above 700~GeV. The spectrum of 
$\tan\beta$ is also rather flat. The full set of the results can be found in \cite{Arbey:2011un}. These results in the unconstrained pMSSM are very different from those obtained in highly constrained scenarios such as mSUGRA.

\section{Implication of Higgs searches}

An alternative way to efficiently constrain SUSY is using the information from the Higgs sector.
In the following, we consider that the new boson discovered at the LHC corresponds to the lightest CP-even Higgs boson. The combination of the Higgs search results presented by ATLAS and CMS are given in Table~\ref{tab:input}.

\begin{table}[t!]
\begin{center}
\begin{tabular}{|c|c|c|}
\hline
Parameter & Value & Experiment \\ \hline \hline
$M_H$     & 125.9$\pm$2.1 GeV & ATLAS \cite{ATLAS:2012gk} + CMS \cite{CMS:2012gu} \\ 
$\mu_{\gamma \gamma}$ & 1.71$\pm$0.33 & ATLAS \cite{ATLAS-2012-091} + CMS \cite{CMS-12-015} \\
$\mu_{Z Z}$ & 0.95$\pm$0.40 & ATLAS \cite{ATLAS-2012-092} + CMS \cite{CMS-12-016} \\
$\mu_{b \bar b}$ & $<$1.64 (95\% C.L.) & CMS \cite{CMS-12-019}\\ \hline
$\mu_{\tau \tau}$ & $<$1.06 (95\% C.L.) & CMS \cite{CMS-12-018}\\ \hline
\end{tabular}
\end{center}
\caption{Input parameters used for the pMSSM study.}
\label{tab:input} 
\end{table}

\begin{figure}[!t]
\begin{center}
\vspace*{0.5cm}
\includegraphics[width=6.7cm]{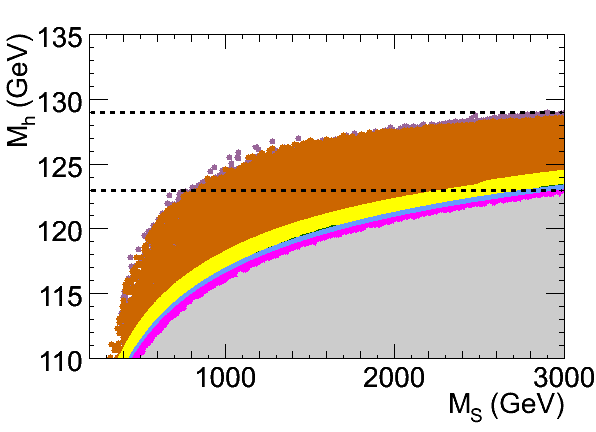}~\raisebox{0.5cm}{\includegraphics[width=2.cm]{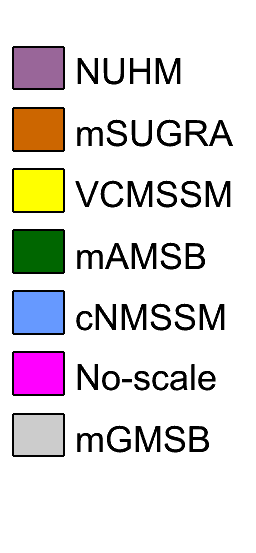}}~\includegraphics[width=6.7cm]{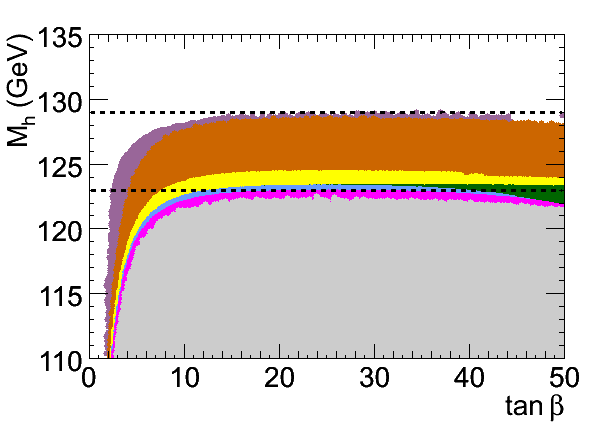}
\end{center}
\caption{The maximal $h$ mass value $M_h^{\rm max}$ as functions of 
$\tan\beta$ (left) and $M_S$ (right) in the mASMB, mGMSB as well as in mSUGRA and 
some of its variants.}
\label{fig:cMSSM}
\end{figure}

In \cite{Arbey:2011ab,Arbey:2012dq}, we have shown that the Higgs mass measurement has strong implications on the constrained MSSM scenarios. This is demonstrated in Fig.~\ref{fig:cMSSM}, where the maximal value of the light Higgs mass is given in mAMSB, mGMSB, mSUGRA and some of its variants, as a function of $\tan\beta$ and the SUSY scale $M_S$. %
The parameters of the models are varied within the ranges given in \cite{Arbey:2012dq}, and the top quark mass is taken to be $m_t=173$ GeV, and $M_S$ is limited to 3 TeV. While mSUGRA and NUHM provide solutions compatible with a Higgs mass $\sim$126 GeV, it is clear that the minimal versions of GMSB and AMSB, and the even more constrained mSUGRA scenarios (VCMSSM, no-scale) are disfavoured.
\begin{figure}[!t]
\begin{center}
\includegraphics[width=7.cm]{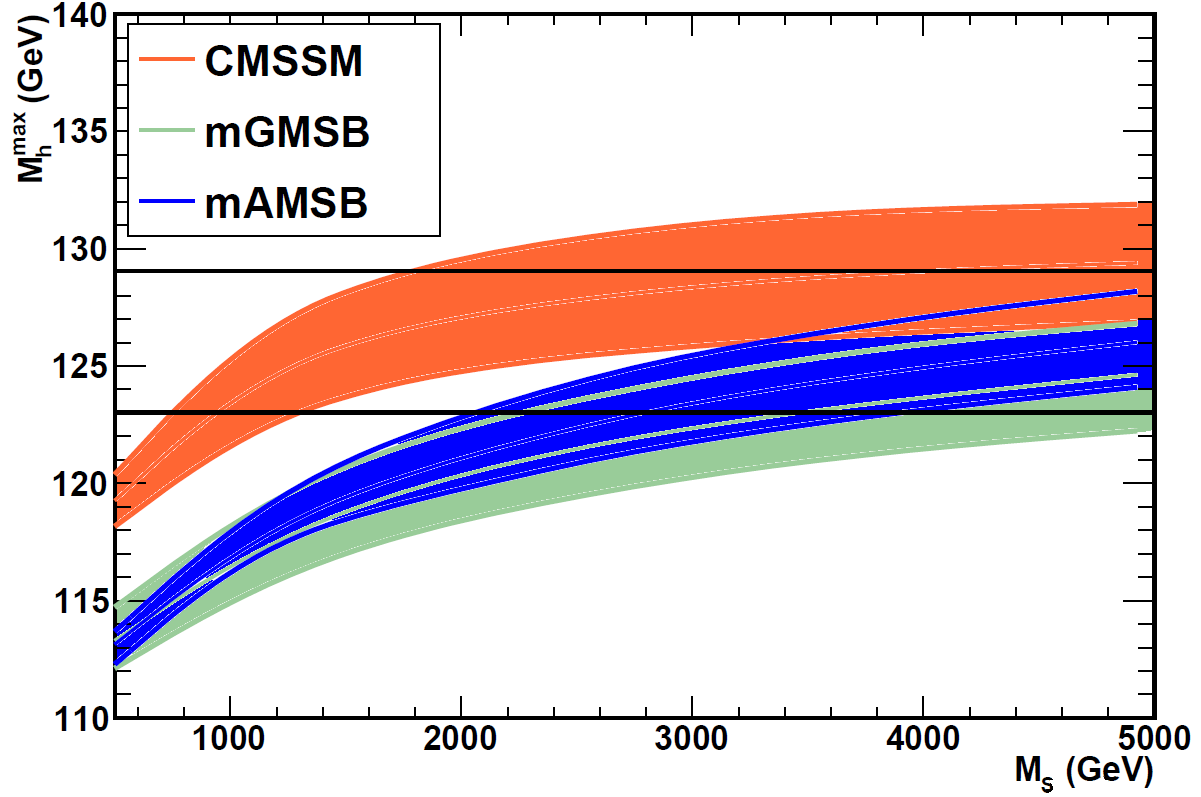}
\end{center}
\caption{Maximal Higgs mass in mSUGRA, mAMSB and mGMSB, as a function of $M_S$ for the top quark mass varied in the range $m_t = 170-176$ GeV.}
\label{fig:top}
\end{figure}

It should be noted that the value of top mass has a significant impact on the maximal Higgs mass, in particular in constrained scenarios, where $m_t$ also enters in the evaluation of
the soft SUSY breaking parameters and the minimisation of the scalar potential. This effect is demonstrated in Fig.~\ref{fig:top} for the minimal SUGRA, AMSB and GMSB models.

\begin{figure}[t!]
\begin{center}
\vspace*{0.5cm}
\hspace*{-0.3cm}\includegraphics[width=0.35\textwidth]{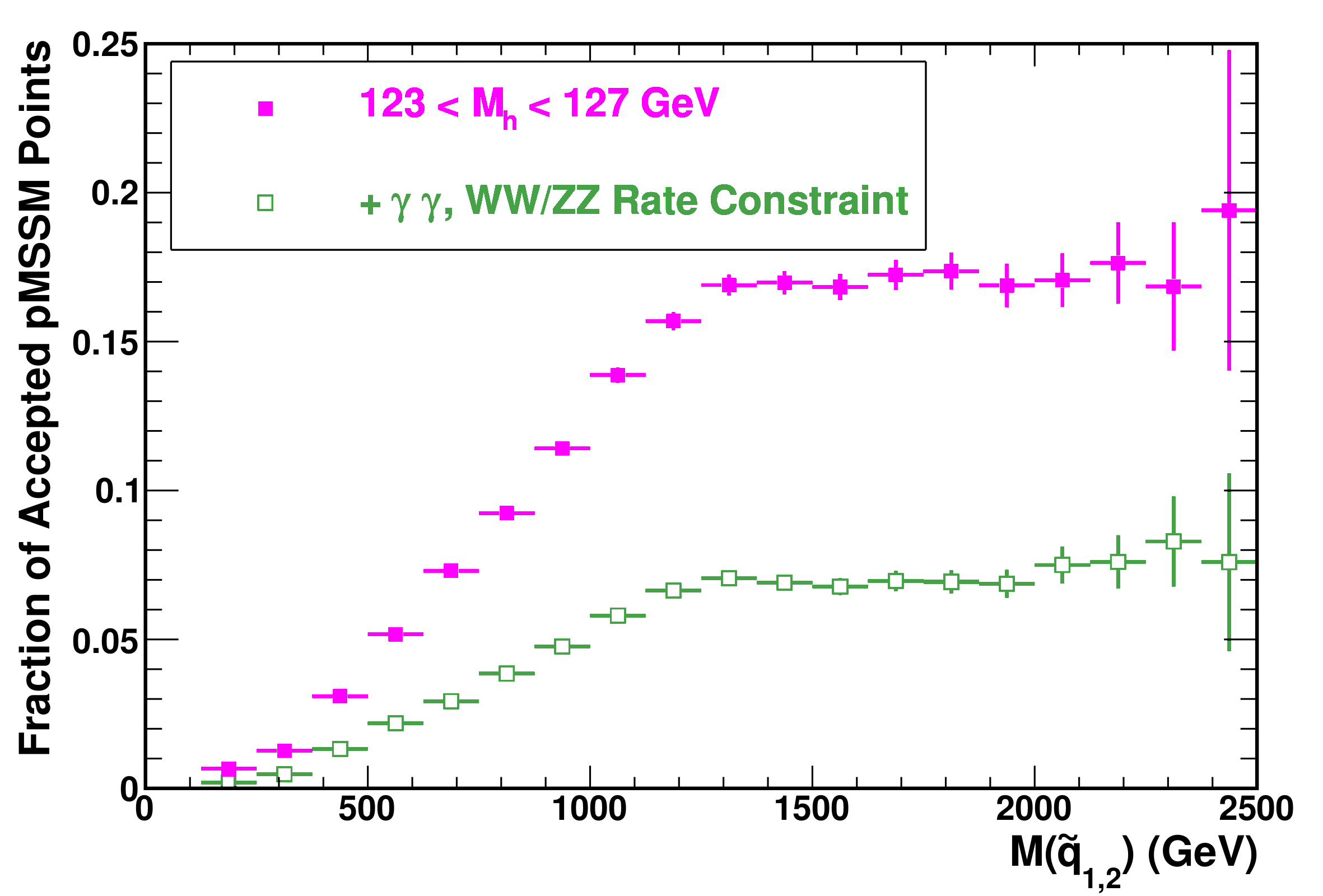}%
\includegraphics[width=0.35\textwidth]{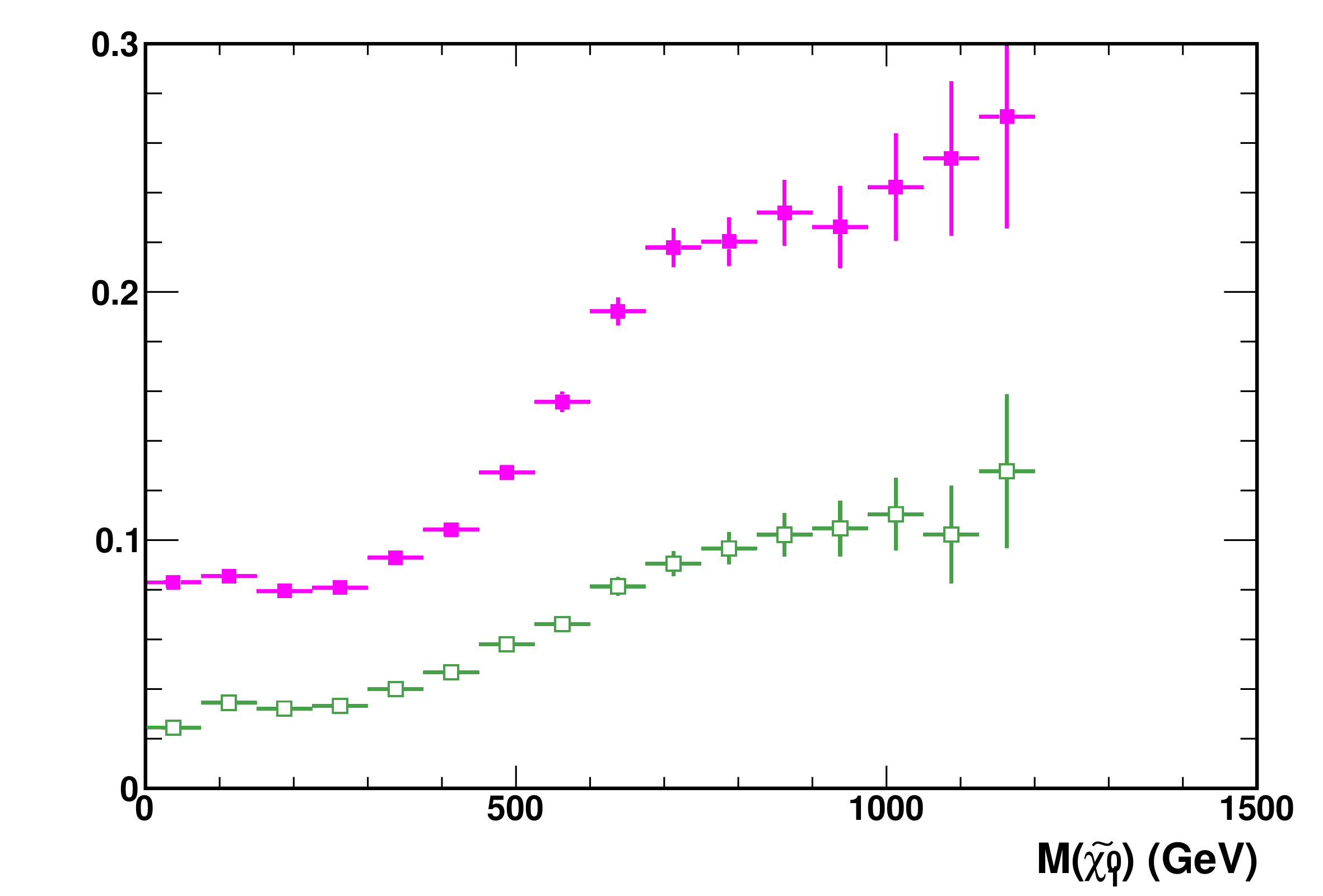}%
\includegraphics[width=0.35\textwidth]{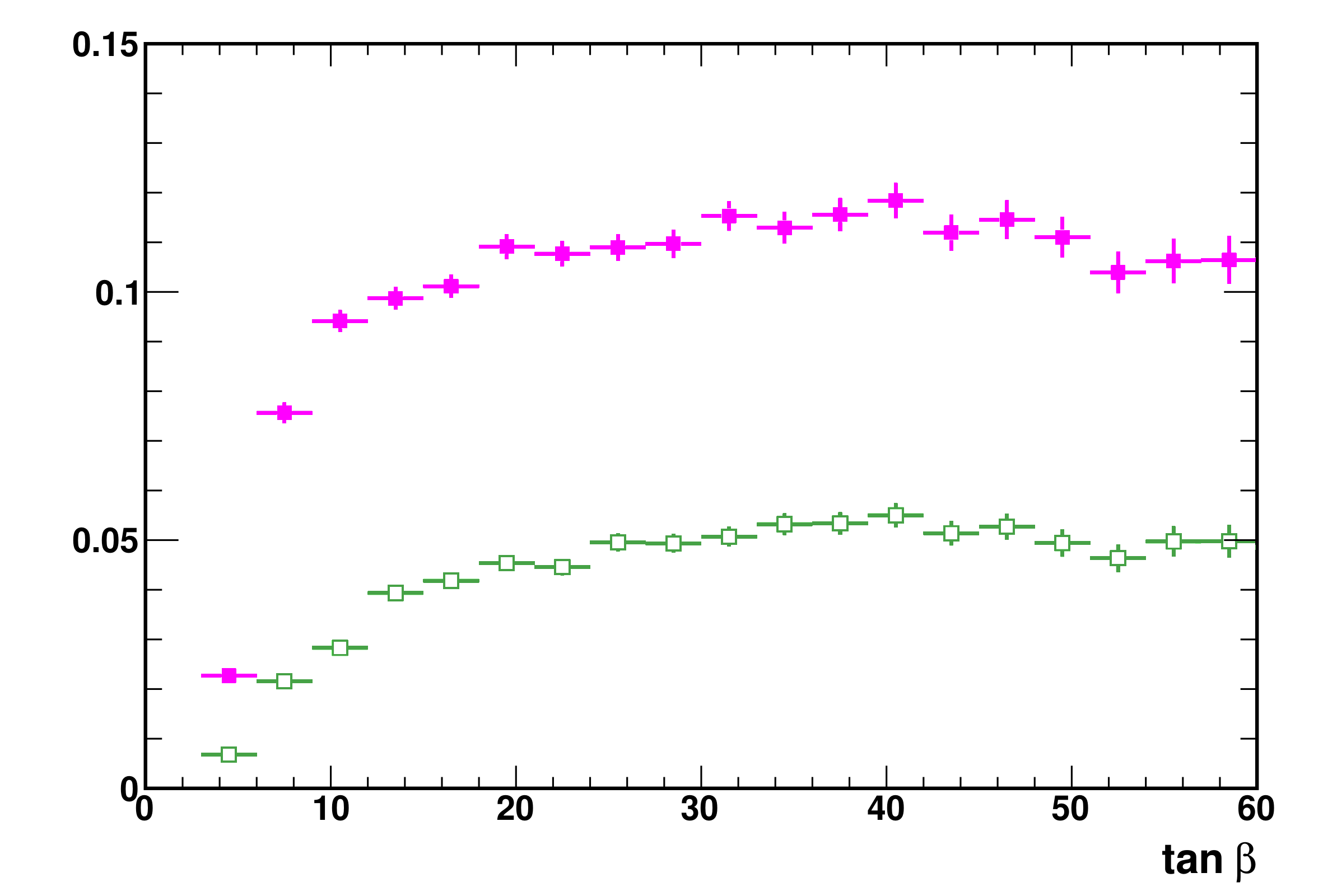}%
\end{center}
\caption{Fraction of accepted pMSSM points, with 123 $< M_{h}<$ 127~GeV (filled squares), 
not excluded by the SUSY searches with 15~fb$^{-1}$ of 7~TeV data as functions of the masses of the lightest squark of the first two generations (left panel), the lightest neutralino $\tilde \chi^0_1$ (central panel) and $\tan \beta$ (right panel). The open square points show the fraction of pMSSM points after imposing the additional requirements on the Higgs rates $\mu_{\gamma\gamma}$ and $\mu_{ZZ}$.\label{fig:msq125}
}
\end{figure}

We turn now to the pMSSM and in Fig.~\ref{fig:msq125} we show the distribution of points compatible with SUSY searches with 15~fb$^{-1}$ of 7~TeV data as well as the Higgs mass constraints and the $\mu_{\gamma\gamma}$ and $\mu_{ZZ}$ signal strengths given in Table~\ref{tab:input}, as functions of the masses of the lightest squark of the first two generations, the lightest neutralino $\tilde \chi^0_1$ and $\tan \beta$.
A comparison with Fig.~\ref{fig:susysearches} reveals first that the Higgs constraints strongly reduce the statistics. The squark and neutralino distribution shapes are quite unaffected, but the small $\tan\beta$ region (below 15) is now more constrained.

\begin{figure}[!t]
\begin{center}
\includegraphics[width=7.5cm]{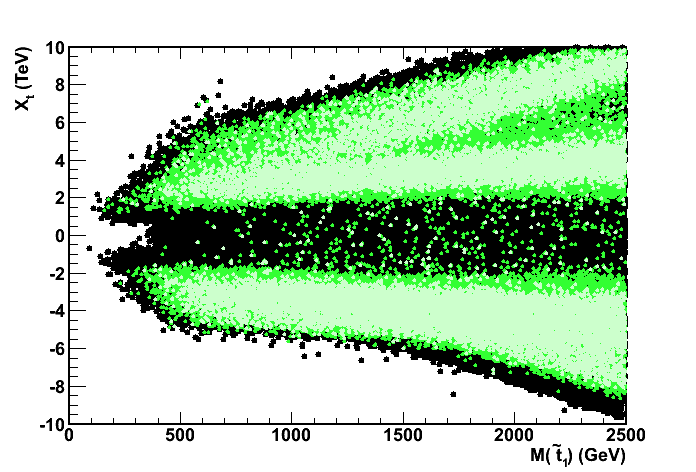}\includegraphics[width=7.5cm]{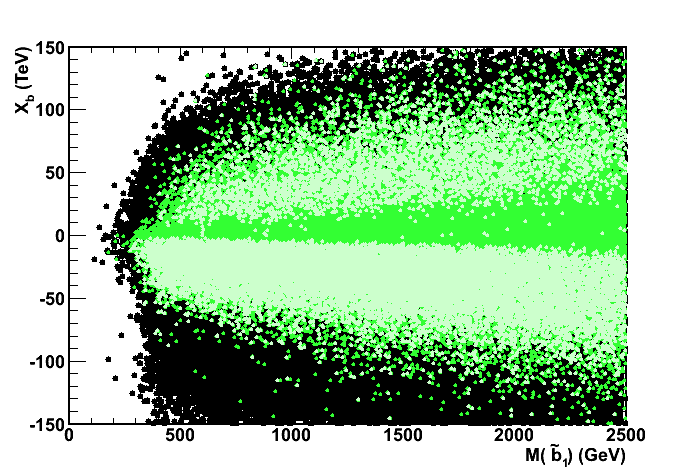}\\
\includegraphics[width=7.5cm]{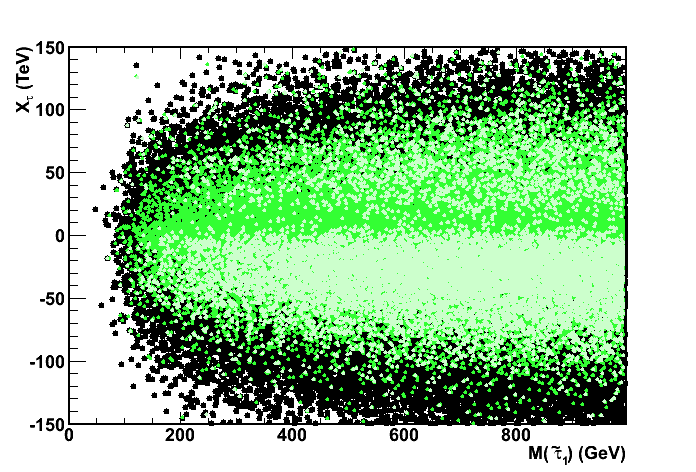}\includegraphics[width=7.5cm]{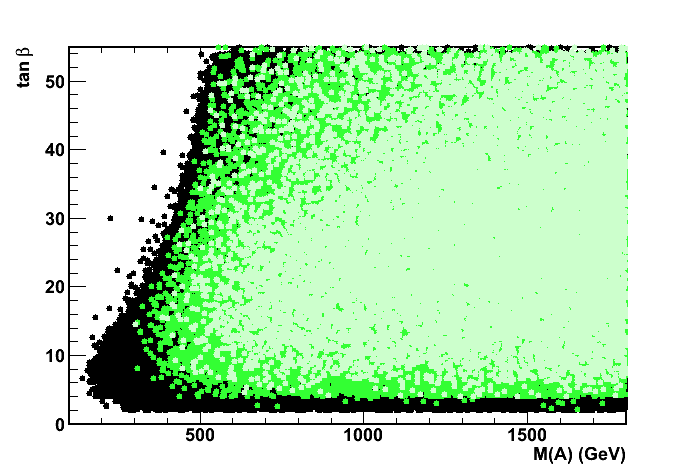}
\end{center}
\caption{Distributions of the pMSSM points in the $(X_t, m_{\tilde t_1})$ (upper left), $(X_b, m_{\tilde b_1})$ (upper right), $(X_\tau, m_{\tilde \tau_1})$ (lower left) and $(M_A$, $\tan \beta)$ (lower right) parameter planes. The black dots show the accepted pMSSM points, those in light (dark) grey the same points compatible at 68\% (90\%) C.L. with the Higgs constraints of Table~\protect\ref{tab:input}.\label{fig:pdf2D}}
\end{figure}

In Fig.~\ref{fig:pdf2D} we present the distribution of pMSSM points compatible with the $h$ boson mass and the observed yields given, in the $(X_t, m_{\tilde t_1})$, $(X_b, m_{\tilde b_1})$, $(X_\tau, m_{\tilde \tau_1})$ and $(M_A$, $\tan \beta)$ parameter planes. To do so, we combined all the constraints in Table~\ref{tab:input} with a $\chi^2$ combination. We notice first that small values of $|X_t|$ are clearly disfavoured, and that stop masses as low as 400 GeV are still compatible with the data. This is mainly the results of the Higgs mass measurement which calls for non minimal mixing in the stop sector. Second, the negative $X_b$ region is favoured by the rate constraints. This can be explained by the fact that a decrease in the $h\to b\bar{b}$ rate, generated in particular by negative $X_b$, would result in an increase of $\mu_{\gamma\gamma}$, which is favoured by present data. Similarly, negative $X_\tau$ are favoured since a reduced $h\to\tau\tau$ rate is more likely to be consistent with the current $\mu_{\tau\tau}$ limit. Finally, $M_A<400$ GeV values are strongly disfavoured by the Higgs mass and rate measurements for any value of $\tan\beta$, and therefore the decoupling regime seems to be favoured by the data.

\section{Conclusion}

We have considered the results from SUSY and Higgs searches at the LHC in the context of the MSSM. Contrary to the constrained MSSM scenarios, in the pMSSM the current LHC limits from the SUSY searches still leave a substantial room for low energy SUSY. This conclusion does not change when we include in addition the information from the Higgs sector, namely the mass and signal yields of the new boson. However, the current Higgs data already point to specific regions of the MSSM, in particular to the decoupling regime with large stop and negative sbottom and stau mixings. With more data becoming available, and more precise experimental results on the Higgs boson properties, more important impacts on the parameter regions of the SUSY scenarios are to be expected.

\end{document}